\documentclass[10pt,letterpaper]{article}
\usepackage{opex3}
\usepackage{color}
\usepackage{threeparttable}
\usepackage{amsmath}
\usepackage{cite}
\usepackage{hyperref}
\usepackage{cleveref}

\usepackage{amssymb}
\usepackage{amsmath}
\usepackage{verbatim}
\usepackage{color}
\usepackage{graphicx}
\usepackage{float}
\usepackage{mathrsfs}
\usepackage{epspdfconversion}

\begin{document}

\title{Double-heralded generation of two-photon-states by spontaneous four-wave-mixing in the presence of noise}
\author{Roger A. Smith, Dileep V. Reddy, Dashiell L.P. Vitullo, and M. G. Raymer}
\date{\today}
\address{Department of Physics and Center for Optical, Molecular, and Quantum Science, University of Oregon, Eugene, OR 97403, United States}
\email{$^*$rsmith13@uoregon.edu}



\begin{abstract}
We present an experimental method for creating and verifying photon-number states created by non-degenerate, third-order nonlinear-optical photon-pair sources. By using spatially multiplexed, thresholding single-photon detectors and inverting a conditional probability matrix, we determine the photon-number probabilities created through heralded spontaneous four-wave-mixing. The deleterious effects of noise photons on reliable heralding are investigated and shown to degrade the conditional preparation of two-photon number states more than they degrade conditional single-photon states. We derive the equivalence between the presence of unwanted noise in the herald channel and loss in the signal channel of heralded experiments. A procedure for characterizing the noise-photon contributions, and a means of estimating the herald noise-free photon-number distribution is demonstrated.
\end{abstract}




\section{Introduction}

Photon-number states can be used for creating photonic operations for quantum information sciences. For example, the teleportation of quantum states is improved by using conditional number-state detection\cite{Opatrny00}. Heralding is one method of conditional number state preparation where the detection of one photon from a photon pair heralds the presence of the other photon. The nonlinear-optical nature of crystalline waveguide structures\cite{Branczyk10}, silicon waveguides\cite{Takesue08,Zhang15}, or silica fibers\cite{Goldschmidt08,McMillan09} are common methods of creating heralded single-photon states and more recently n-photon states\cite{Achilles06,Sullivan08}. $\chi^{(3)}$ media are advantageous over $\chi^{(2)}$ waveguide structures for creating sources easily integrable into fiber networks for quantum communication\cite{Li04,Sharping06}, but they tend to suffer from higher levels of unwanted noise photons created by processes such as Raman scattering. Thus, it is important to devise ways to characterize and counteract this noise.

	Time-correlated photon pairs are created via spontaneous four-wave-mixing (SFWM) in a third-order nonlinear optical medium. When one photon of the pair is detected in the herald channel, then the probability of detecting the paired signal photon approaches one because of the time-correlated nature of the photon pair. Any spurious photons in the herald or signal channel can degrade the fidelity of the number states created by heralded SFWM. Inelastic Raman scattering in the fiber produces unwanted photons in a broad spectrum mostly downshifted in frequency from the pump. In silica fibers, the scattered Raman spectrum peaks near $13.2$ THz below the pump frequency. The creation of a photon pair and a noise photon are independent processes. Suppression of unwanted Raman scattering has been shown in fibers with large birefringence\cite{BJSmith09}, through orthogonal phasematching\cite{Lin06,Lin07}, or by cooling the optical fiber\cite{Takesue05,Lee06}. Additionally, any photons from other sources such as the pump laser that leak into the herald or signal channel will degrade the photon statistics.
	
	In the low-gain regime, the probability, $\epsilon_{Pair}$, of creating a photon pair within a time window, i.e. a signal and idler photon via SFWM during one pump pulse, is very small and it may be on the same order as the probability of creating a noise photon in the herald channel, $\epsilon_{Noise}$. In this regime, a single-heralding detection scheme overcomes the effects of noise photons in the herald channel.  For example, if $\epsilon_{Pair} = \epsilon_{Noise} = 1/100$ then conditioned on detecting a herald event, either from an idler or noise photon, the probability for a signal photon to be present is $\epsilon_{Pair}/\left(\epsilon_{Pair} + \epsilon_{Noise}\right) = 0.5$. This result suggests that noisy heralding masquerades as a loss in the signal channel. In many applications, such as second-order coherence, or $g^{(2)}$, measurements, this only has the effect of decreasing the signal count rate. 

	In contrast, double heralding for two-photon generation is more susceptible to false herald events caused by noise. Since pair creation events are independent, the probability of creating two pairs of photons is proportional to the square of the probability of creating one photon pair, i.e., $\epsilon_{2Pair}\sim\epsilon_{Pair}^2$. In the example above, the probability of creating one photon pair and one noise photon, which is given by $\epsilon_{Pair}\times\epsilon_{Noise}$, or of creating two noise photons in the herald, which is given by $\epsilon_{Noise}^2$, are the same value as $\epsilon_{2Pair}$. This independence of pair creation events allows for a larger influence from noise photons in double-heralding detection schemes than in single-heralding schemes. Using the same numbers as above, the probability for two signal photons conditioned on the detection of two herald photons is $\epsilon_{Pair}^2/\epsilon_{2Heralds}=0.25$, where the denominator is given by $\epsilon_{2Heralds}=\epsilon_{Pair}^2+\epsilon_{Noise}^2+2\left(\epsilon_{Pair}\times\epsilon_{Noise}\right)$. This reduction of heralded pair generation not only reduces the count rate of the two-photon component, that is, acts effectively as a loss of signal photons, but also introduces unwanted single-photon components. Using the numbers in the example above gives a conditioned probability for a single signal photon equal to $0.5$, which dominates the two-photon component. Again the false heralding acts as an effective loss, as we show in \Cref{sec:noisyheralding}.

	In this study, we demonstrate that heralding two-photon states is far more susceptible to corruption by noise heralds, and we show a way to analyze and in a sense overcome this corruption. We present a method for measuring the number statistics of light created via SFWM in the presence of unwanted noise herald photons. Our technique can separate the contributions due to the targeted SFWM process and independent noise photons by measuring the ``Klyshko" efficiencies of detecting coincidences from time-correlated photon pairs. The Klyshko heralding efficiency is the conditional probability that given a herald detection a signal photon will also be detected.The Klyshko heralding efficiencies provide valuable information about the photon number distributions of light that would be created if the noise photons from Raman scattering or pump photons were not present in the herald beam. With this technique, one can decide the benefits of additional means for mitigating noise photons and determine the necessity to do so.
\section{Spatially-Multiplexed Photon Detection}
\begin{figure}[h!]
	\centering
	\includegraphics[width=0.75\textwidth]{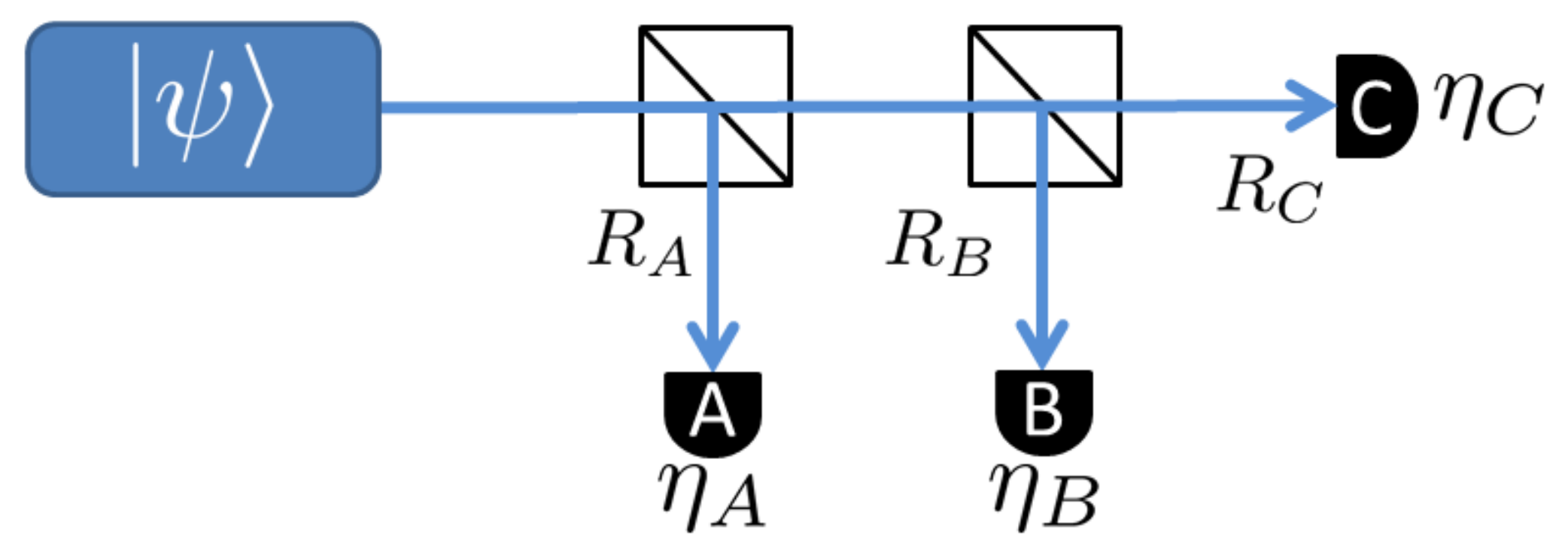}
	\caption{Spatially multiplexed detector setup with three detectors to approximate a number-resolving detector. Each detector receives a fraction $R_i$ of the light from the original state and has an efficiency of $\eta_i$ of detecting an incident photon.}
	\label{fig:spmultdet}
\end{figure}

	In the absence of number-resolving single-photon detectors, thresholding detectors multiplexed together approximate number-resolving detection capabilities by dividing a signal in time or space\cite{Fitch03,Rehacek03,Sperling12,Chrapkiewicz14,Rohde07,Avenhaus10,Bartley13}. Thresholding detectors create a ``click" if one or more photons is detected. This approach is valid when the number of detectors is larger than the number of photons under measurement. Spatially multiplexed thresholding detectors, typically avalanche photodiodes or APDs, can be used with high-repetition-rate sources, limited only by the dead time of the detectors. An example of three spatially multiplexed detectors is shown in Fig.\ref{fig:spmultdet}. 

	The non-unity efficiency and thresholding nature of the detectors affects the probability of detection if $n$ photons are incident. The probabilities of detection events are given by the conditional probability matrix multiplied by the photon-number distribution\cite{Achilles03,Achilles04,Banaszek03}. For example, for three thresholding detectors, the probabilities that one, two, or three detectors (regardless of which) ``fires" or ``clicks" is given by:
\begin{align}
\begin{pmatrix} q_1 \\ q_2 \\ q_3
\end{pmatrix} =
  \begin{pmatrix}
   P(1|1) & P(1|2) & P(1|3) \\
   0 & P(2|2) & P(2|3) \\
   0 & 0 & P(3|3) 
  \end{pmatrix}
  \begin{pmatrix} p_1 \\ p_2 \\ p_3\\
\end{pmatrix} 
\label{eq:pcmatrix}
\end{align}
	where $q_k$ is the probability of $k$ clicks, $p_n$ is the probability of $n$ photons incident upon the system of spatially multiplexed detectors, and the matrix elements $P(k|n)$ are the probability to produce $k$ clicks, given $n$ incident photons. Here we assume that the probabilities for creating photon pairs larger than $n=3$ are negligibly small, and truncate the conditional probability matrix at three photons. This method easily generalizes to higher photon numbers.

	The conditional probability matrix elements are calculated using the efficiency of detecting a photon for each detector. The matrix elements are shown explicitly in Appendix A. Set theory is used to combine the single-, double-, and triple-detection events recorded to determine the values for $q_k$.

The conditional probability matrix could be expanded to include a row for the probability of zero detectors clicking, $q_0$. However, the equation for $q_0$ in terms of $\{p_0,p_1,p_2,p_3\}$ would not be linearly independent of the other rows in the matrix. Instead, we calculate the probability of zero clicks by using the assumptions that the probability of photon pairs larger than $3$ is zero and that the sum of probabilities of all possible number of photon pairs is 1: 
\begin{align}
p_0=1-(p_1+p_2+p_3)
\label{eq:p0}
\end{align}

	The conditional probability matrix is an upper triangular matrix that maps $n$ incident photons into $k$ detection events.The equations for the probabilities of $k$ detection events are linear in the photon-number probabilities, and therefore direct inversion of Eq. \ref{eq:pcmatrix} is possible to yield the photon-number probabilities. Note that the linear inversion method does not constrain the probabilities to be positive in the presence of statistical or systemic error.

	In our experiment, we are detecting number states and not explicitly Fock states, which would imply single-mode excitation only. This method applies either to single-mode Fock states or multimode number states, in contrast to some other methods, such as optical homodyne tomography\cite{Munroe95,Lvovsky01,Cooper13,Ourjoumtsev06}. Tomographic methods are not easily implementable for characterizing multimode number states, as those methods detect a single mode for a given local oscillator temporal shape. In our scheme, the conditional probability matrix maps input photon-number distributions to a distribution of detection events, regardless of the number of spatial or temporal modes that the photons occupy. 
%
%
%
%

\section{Effects of Noisy Heralding}
\label{sec:noisyheralding}
	A targeted nonlinear source process produces a joint probability distribution of photons in the signal (subscript $s$) and idler (subscript $r$) channels denoted by $\bar{P}_{rs}(n_s,n_r)$, where the bar denotes the absence of noise in the herald (idler) channel. This definition is general and makes no assumptions about the joint probability distribution of the source. (For example, if the source was a perfect, noise-free SFWM source, the distribution would be $\bar{P}_{rs}(n_s,n_r) = \delta_{n_rn_s}\bar{P}(n_s)$.) We only assume that all photons in the herald channel arise from the targeted source process. Given the detection of a photon in the idler channel, the conditional probability that a photon exists in the signal channel is:
\begin{align}
\bar{P}_{sr}(n_s|n_r) = \frac{\bar{P}_{rs}(n_s,n_r)}{\sum_{m=0}^\infty \bar{P}_{rs}(m,n_r)} = \frac{\bar{P}_{rs}(n_s,n_r)}{\bar{P}_r(n_r)}
\label{eq:noiselesssig}
\end{align}
	where $\bar{P}_r$ denotes the noise-free probability distribution in the idler channel.

	Now consider the effects of added, uncorrelated noise in the herald channel. Detection events in the herald channel result from either a photon created by the nonlinear source in the targeted process or by some other independent process. Suppose the probability of a given click in the herald channel is from the target process is $\eta_r$. Then the probability that a given click arises from a noise process is $1-\eta_r$.  Therefore, the conditional probability that given $k$ detection events in the idler channel, $m$ of those are from the target process is a binomial distribution:
\begin{align}
p(m|k) = \binom{k}{m}\eta_r^m(1-\eta_r)^{k-m}
\label{eq:noisebinom}
\end{align}

	The conditional probability of detecting $n_s$ photons in the signal channel given $k_r$ herald events is then the convolution of the targeted nonlinear process joint probability distribution and the noise distribution:
\begin{align}
P(n_s|k_r) =\sum_{m=0}^{k_r}\bar{P}_{sr}(n_s|m) p(m|k_r)=\sum_{m=0}^{k_r}\binom{k_r}{m}\eta_r^{m}\left(1-\eta_r\right)^{k_r-m}\bar{P}_{sr}(n_s|m)
\label{eq:pnkherald}
\end{align}
	The form of Eq. \ref{eq:pnkherald} is precisely that resulting from a fictitious beamsplitter with probability of transmission being the conditional probability that a given herald detection event resulted from a herald photon of the target process, $\eta_r$.
	
	That is, if we were to assume there is no false heralding, but rather some of the photons created in the signal channel via the targeted nonlinear process are lost, then the probability for $n$ signal photon detections when there are $k$ photons present is $\left(n\leq k\right)$:
\begin{align}
	P_S(n|k)=\sum_{m=1}^{n}\binom{k}{m} \left(1-\rho_{Loss}\right)^m\left(\rho_{Loss}\right)^{k-m}\bar{P}_{sr}(m|k)
	\label{eq:signaleta}
\end{align}
	where $\rho_{Loss}$ is the amount of the signal channel that is lost. For example, this can be thought of as a beam splitter in the path of the signal channel with a reflection coefficient of $\rho_{Loss}$. 	Indeed, this equation has the same form as Eq. \ref{eq:pnkherald} with $\eta_r = (1-\rho_{Loss})$. Therefore, we have shown that heralding in the presence of noise is the same as noise-free heralding with a lossy signal channel in the case where the processes of photon-pair creation and noise creation are independent.

	As a result, the photon distribution in the signal channel, conditioned upon a noise-free herald channel, can be inferred by introducing a fictitious loss into the conditional probability matrix inversion calculation. The loss is introduced as a decreased efficiency of detection in the elements of the conditional probability matrix in Eq. \ref{eq:pcmatrix}, as described in more detail below.
%
%
%

\section{Experimental Setup}
\begin{figure}[h!]
	\centering
	\includegraphics[width=0.85\textwidth]{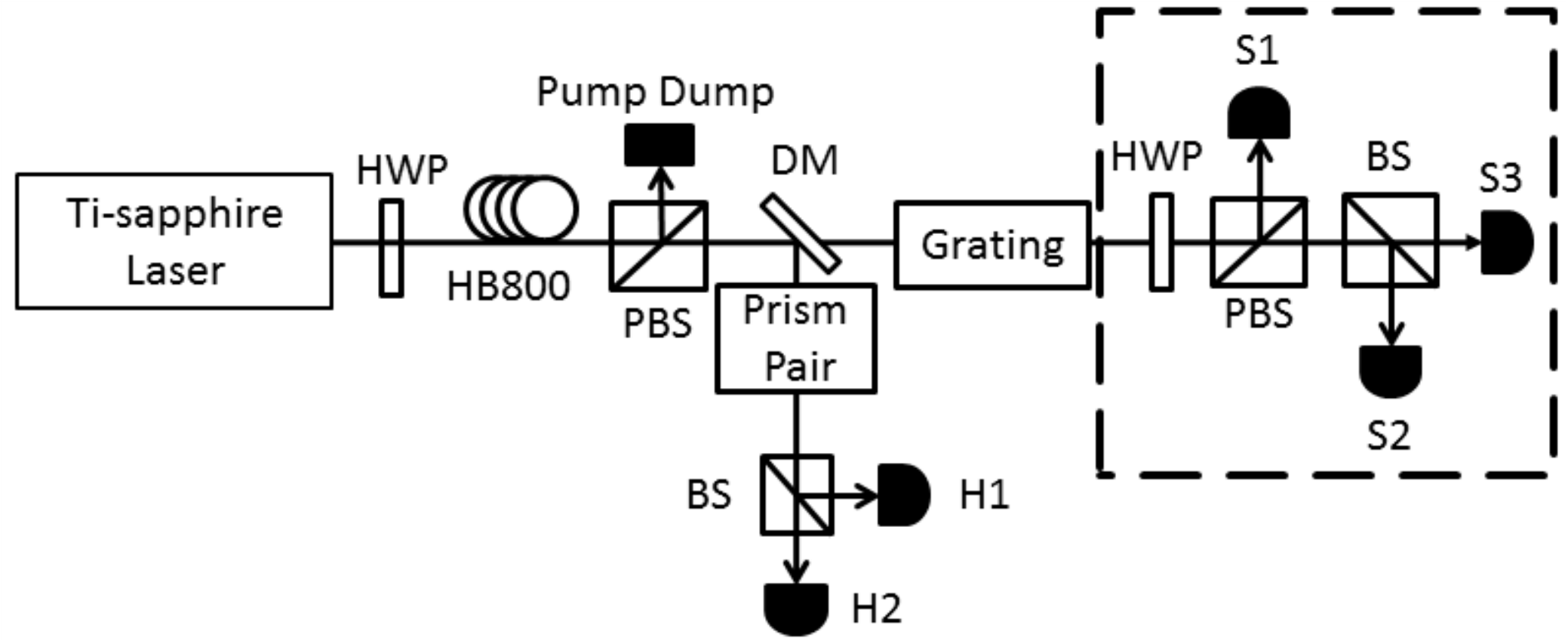}
	\caption{Experimental setup for measuring conditionally prepared number states. The multiplexed detectors are located within the dashed box. The two heralding detectors and the three signal detectors are connected to an FPGA and a computer for storing data. HWP denotes a halfwave plate; DM denotes a dichroic mirror; BS denotes a beamsplitter; PBS denotes a polarizing beamsplitter.}
	\label{fig:exsetup}
\end{figure}

	The experimental setup is shown in Fig. \ref{fig:exsetup}. A single-mode, birefringent optical fiber (Fibercore HB800) with length of $20$ m is pumped along the slow axis by a pulsed, mode-locked titanium-sapphire laser operating at $802$ nm, with a transform-limited pulse duration of $36$ ps, and repetition frequency of $76$ MHz. For each pair of pump photons that are spontaneously annihilated in the fiber, a non-degenerate, time-correlated photon pair is created polarized along the fast axis of the fiber. The frequency-upshifted photon is denoted as the signal photon with a central wavelength of $686$ nm and the frequency downshifted photon is denoted as the idler photon with a central wavelength of $965$ nm and is used to herald the signal photon. The SFWM photons are separated from the pump by $437$ THz, far beyond the peak of the Raman spectrum at $13.2$ THz below the central pump frequency.

After the fiber, the pump light reflects off a polarizing beamsplitter (PBS) that transmits the SFWM photon pair. The herald beam reflects off a dichroic mirror (DM), transmits through a prism pair and iris, a narrowband filter (FWHM $\sim 2$ nm), and is split at a $50/50$ beamsplitter (BS) to two multimode fiber-coupled avalanche photodiodes (APD). The APD’s have a quantum efficiency of approximately $15\%$ at $965$ nm and the coupling efficiency is estimated at $80\%$ through the use of alignment lasers close to the central wavelength of the herald photons. 

	The signal beam transmits through the dichroic mirror and reflects off a blazed grating ($2200$ lines/mm) before the spatially multiplexed detectors. A halfwave plate (HWP) is oriented such that $30\%$ of the incident light is reflected at a PBS and coupled to an APD through a multimode fiber. The transmitted light is split at the $50/50$ BS to two fiber-coupled APDs. The APDs have a quantum efficiency of approximately $60\%$ at the signal wavelength. Using a tracer beam close to the signal wavelength, the overall efficiency in the signal arm is estimated to be $36\%\pm5\%$. This efficiency includes optical transmission losses from elements between the fiber and the APDs, coupling losses into the multimode fibers, and the quantum efficiency of the APDs. The signal arm efficiency and the beamsplitter ratios are important parameters that will be used in the inversion algorithm (see Eqs. \ref{eq:PAclick}-\ref{eq:PABCclick}) to estimate the signal-photon-number distributions.

	The APDs operate in Geiger mode, and so act as thresholding detectors, which generate a `click' if one or more photons is detected. A field-programmable gate array (FPGA) collects the electronic signals from the detectors and records all singles and combinations of coincidences. The data is stored on a computer through LabView. By combining the singles, two-fold, and three-fold coincidences, we are able to extract the number of detections where one and only one, two and only two, or three detectors fired. As a result, we are able to collect data at a rate of a few MHz without having to deal with detector dead-time issues. The increase in speed is a major benefit to using spatially-multiplexed detectors over temporally-multiplexed detectors.
%

%
%
%

\section{Results and Analysis}
\label{sec:results}

\begin{figure}[H]
	\centering
	\includegraphics[width=\textwidth]{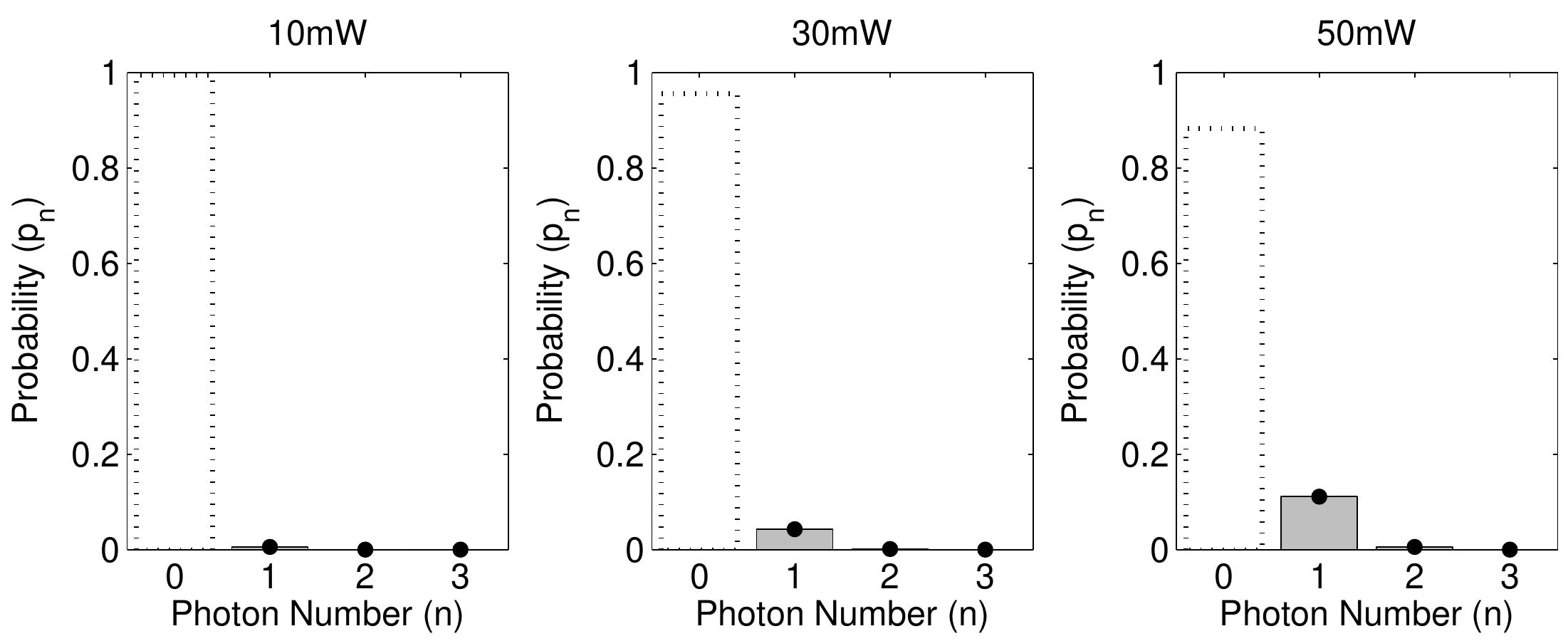}
	\caption{Photon-number probabilities measured without conditional detection of herald photons. The vacuum component is outlined with a dashed line. Error bars are present and are smaller than the markers.}
	\label{fig:pn0hsyseff}
\end{figure}

The analysis is performed for detection events conditioned on detection by zero, one, or both herald detectors for average pump powers between $10$ mW to $50$ mW ($3.6$ W to $18$ W peak power). For the initial experiments, the fiber had length $20$ m. The data collected without heralding is shown in Fig. \ref{fig:pn0hsyseff}. All statistical error bars are computed by propagation of errors through the linear-inversion algorithm. The measured distributions contain a large vacuum component and confirm that the fiber is pumped in the low-conversion regime. 

%
%

\begin{figure}[H]
	\centering
	\includegraphics[width=\textwidth]{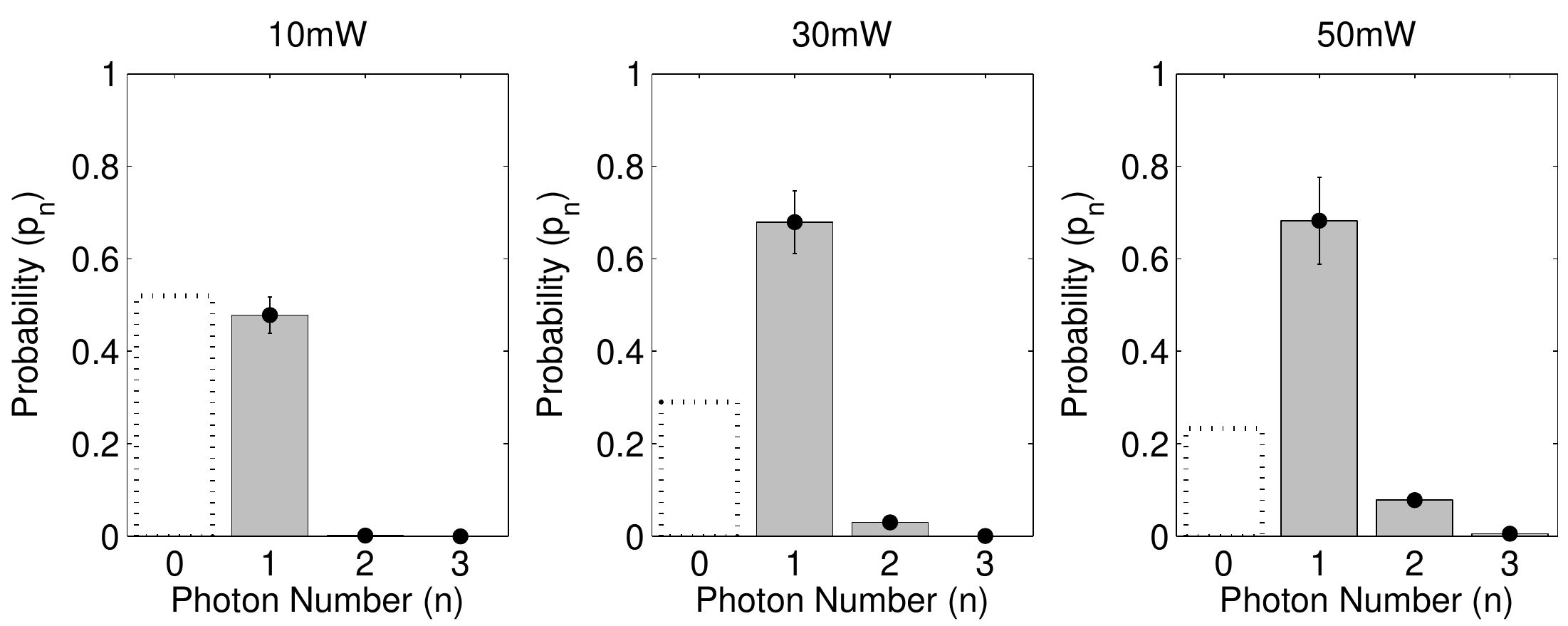}
	\caption{Photon-number probabilities measured with single herald conditional preparation. The vacuum component is suppressed while the $p_1$ component is high, showing a dominant one-photon-number component.}
	\label{fig:pn1hsyseff}
\end{figure}

\begin{figure}[H]
	\centering
	\includegraphics[width=\textwidth]{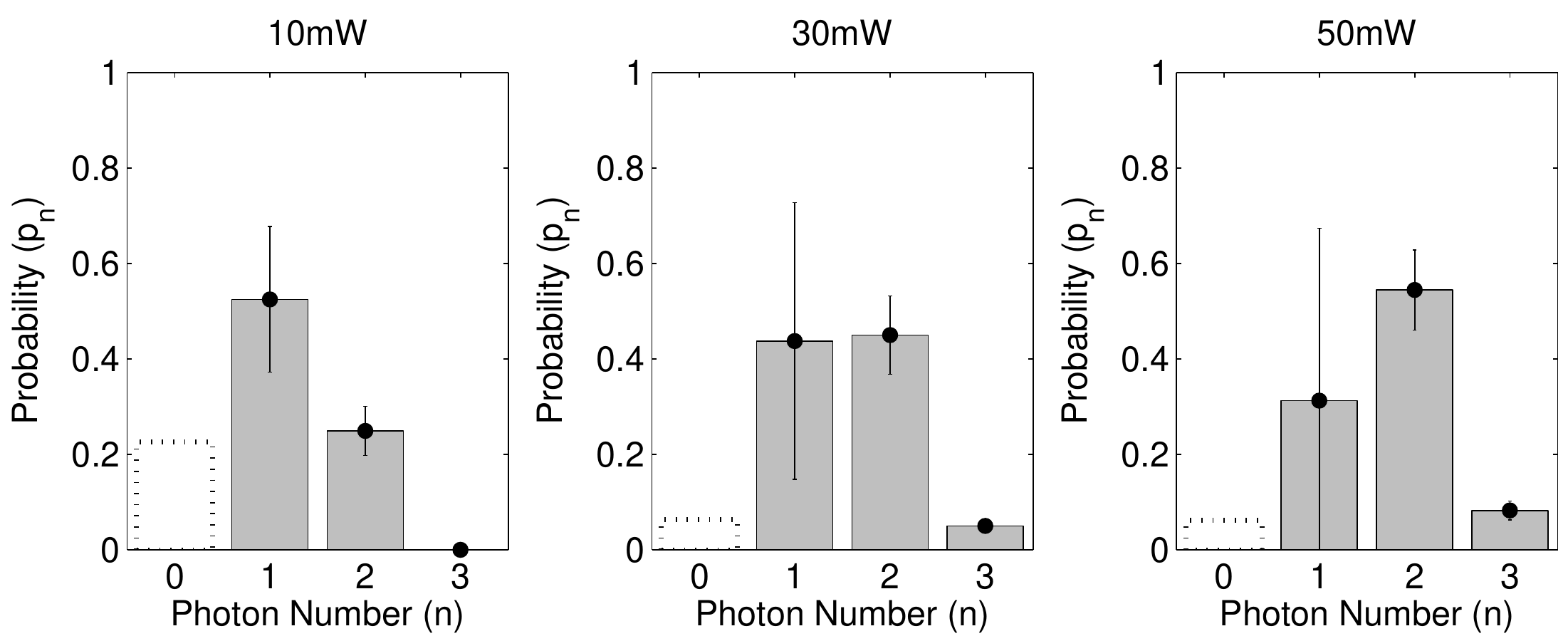}
	\caption{Photon-number probabilities of double-heralded states. The one-photon contribution is large relative to the two-photon contribution, in contrast to the desired outcome of a pure two-photon state.}
	\label{fig:pn2hsyseff}
\end{figure}

Conditioning detection events on a single herald suppresses the vacuum component and increases the single-photon component dramatically, as is well known. The constructed distributions are shown in Fig. \ref{fig:pn1hsyseff}. Conditional preparation of number states with a single herald is an effective method of suppressing noise photons such as spontaneous Raman scattered photons or leaked pump photons when the experiment is insensitive to vacuum contributions, such as second-order coherence, or $g^{(2)}(\tau)$, measurements.
%
%
	After conditionally preparing the states with a double-heralded detection event, a strong one-photon contribution remains, shown in Fig. \ref{fig:pn2hsyseff}. The two-photon contribution increases, but is not indistinguishable from the one-photon contribution due to systematic uncertainty. This is a clear example of the pitfall with double heralding described in the introduction. The probability to create two pairs is of the same order as the probability to create one SFWM photon pair and one noise photon in the heralding channel. In the low-gain regime, pair creation events are independent and the probability to create two pairs is the probability of creating one photon pair squared. The large error bars are primarily attributed the systematic error in the experiment, specifically the uncertainty in the overall efficiency in the signal channels, including coupling efficiency, transmission losses, and quantum efficiencies.

	If the probability, $\delta$, of creating a noise photon (through Raman scattering or leakage from the pump) is on the same order as creating a single pair of photons, then a double heralding experiment will be highly susceptible to false double-herald events. In this case, the probability of creating a noise photon and a single photon pair scales as $\delta^2$, which is on the same order as the probability of producing two pairs of photons. This exposes an issue with heralding experiments conditioned on photon numbers larger than one. 

	In the low-gain limit, for single-heralded sources, noise in the herald channel effectively adds a vacuum component to the heralded-signal number distribution, thus approximating a low-efficiency or low-count-rate single-photon source. In double-heralding sources, however, noise heralds pollute the signal-number distribution with a significant one-photon component, whose effects in subsequent experiments, such as $g^{(2)}$ measurements, can be very different from a mere addition of vacuum.

%
%

\section{Efficiency of Measuring Coincidences}

Although double-heralded conditional preparation is susceptible to noise photons in the herald channel, we show how to reconstruct the underlying true signal-photon distribution that would be observed in the absence of noise in the herald channel. This relies on the noise being statistically independent from the target process. By utilizing the temporal correlation of the creation of photon pairs, we can define an efficiency of detection of correlated photons, which is called the Klyshko heralding efficiency\cite{Achilles04}. Klyshko proposed a method for calibrating the efficiency of a detector with correlated photon pairs that has been studied in depth by Migdall and others\cite{Klyshko80,Ware04,Polyakov07}. When using only one detector each in the signal and herald channels, the Klyshko heralding efficiency for the signal detector is defined as:
\begin{align}
	\eta_{K} = \frac{P(H,S)}{P(H)}
\label{eq:kpeff}
\end{align}
	where $P(H,S)$ is the probability of detecting a coincidence between detectors $S$ and $H$ and $P(H)$ is the probability of a herald detection. Therefore, the Klyshko heralding efficiency is the conditional probability that given a herald detection, a signal photon will also be detected. If the signal and herald channels are free of noise photons, then in a long counting window, the number of coincidence detections and herald detections are, respectively:
\begin{align}
	N_{HS} = \eta_H \eta_S \gamma \\
	N_H = \eta_H \gamma
\label{eq:coincount}
\end{align}
	where $\eta_H$ is the detection efficiency of the herald channel (including transmission losses and detector efficiency), $\eta_S$ is the detection efficiency of the signal channel, and $\gamma$ is the number of photon pairs produced within the same time interval. Without noise in the herald or signal channel, the Klyshko heralding efficiency in some long counting time interval reduces to the efficiency in the signal channel:
\begin{align}
\eta_{K}= \frac{P(H,S)}{P(H)} = \frac{N_{HS}}{N_H} = \frac{\eta_S\eta_H \gamma}{\eta_H \gamma} = \eta_S
\label{eq:keff}
\end{align}
	We assume that noise photons are present only in the herald channel. Cross-polarized phasematching has been shown to cause negligible amounts of upshifted Raman scattered photons\cite{Lin07}, consistent with our observations. If there are noise photons in the herald channel, the numerator will be unchanged and the denominator will increase, decreasing the Klyshko heralding efficiency. If the noise is independent from the pair production process, the noise can be added to the number of herald detections in the denominator:
\begin{align}
	\eta_K =\frac{\eta_S \eta_H \gamma}{\eta_H \gamma + \eta_H \sigma}= \eta_S \frac{\gamma}{\gamma+\sigma}
\end{align}
	where $\sigma$ is the number of noise photons in the herald channel. The Klyshko heralding efficiency accounts for false herald events by scaling the targeted process of photon-pair creation by the number of herald detection events, both from photon-pairs and noise.

	In the herald channel, the number of spontaneous Raman photons depends linearly on pump power, $P$ and fiber length, $L$, $N_R \propto PL$  while SFWM pair creation scales as $N_{SFWM} \propto P^2L$, when the pump is pulsed with a pulse duration shorter than the travel time through the optical fiber and if there is no spectral filtering within the phasematching bandwidth of the fiber\cite{Lin06}. Pump photons that leak into the herald channel are independent of fiber length and depend linearly on pump power. Exponential length-dependent fiber losses are assumed to be negligible. Thus, we assume that noise photons are present only in the herald channel, consistent with our observations. Additionally, the dichroic mirror used to separate the herald and signal beams reflects the pump wavelength. The ratio of the number of SFWM photons to the sum of SFWM and noise photons provides a power- and length-dependent factor that scales the Klyshko heralding efficiency:

\begin{align}
\eta_{K}=\eta_{S}\times\frac{P^2L}{P^2L+\beta PL+\alpha P}
\label{eq:nkpscale}
\end{align}

	where $\beta$ is the power at which Raman scattering matches the SFWM gneration rate, $\alpha$ is defined similarly for leaked pump photons, and $\eta_{S}$ is the total (source independent) system detection efficiency in the signal channel. $\alpha$ has units of power times length.

\begin{figure}[h!]
	\centering
	\includegraphics[width=.85\textwidth]{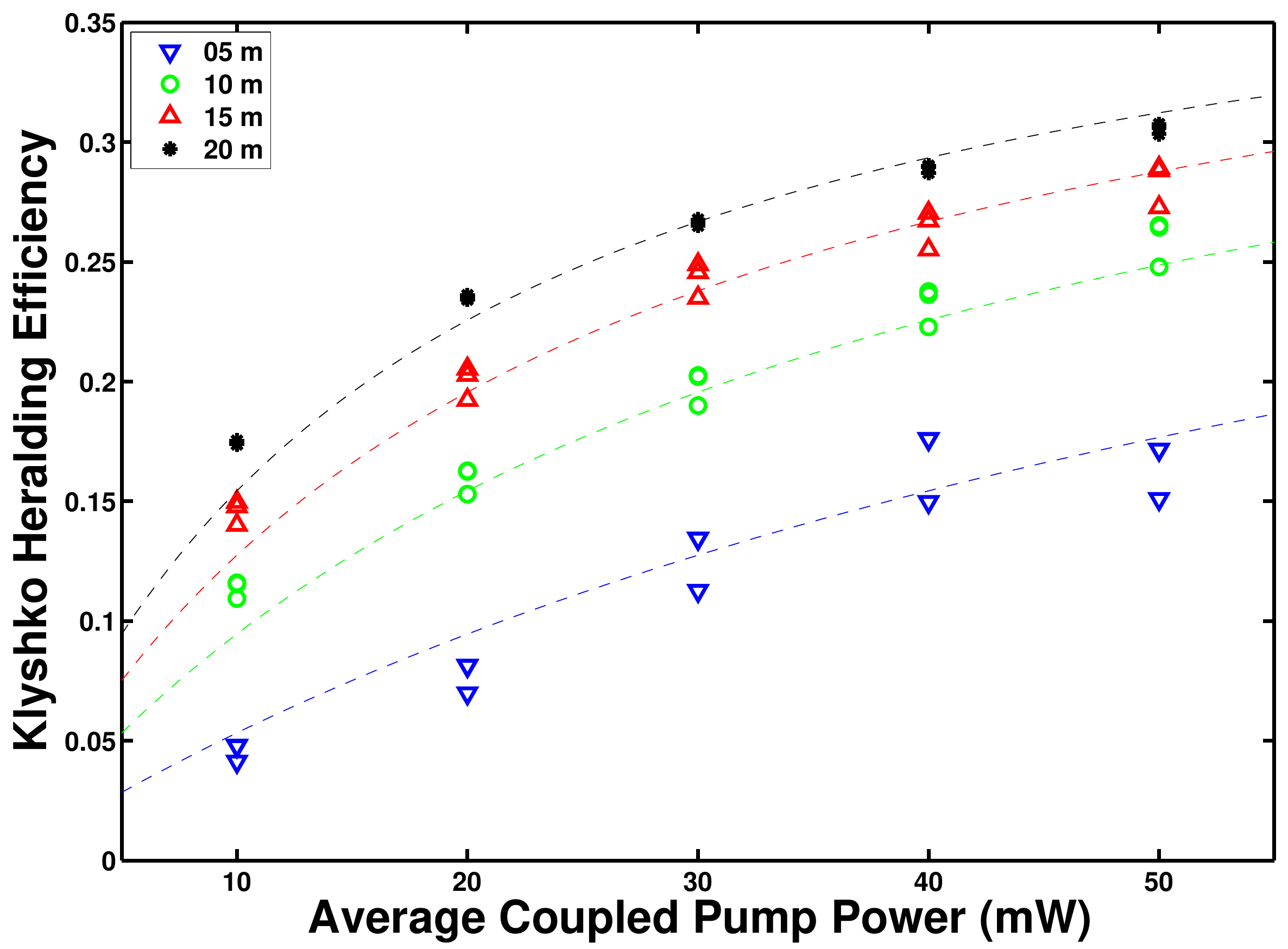}
	\caption{Meausred Klyshko heralding efficiencies for different average pump powers and fiber lengths with the fits to Eq. \ref{eq:nkpscale} denoted by the dashed lines. There are three points presented for each pump power and fiber length representing the Klyshko heralding efficiency for each signal detector. Some points overlap where values are nearly identical.}
\label{fig:exkeff}
\end{figure}

	For our experiment, Eq. \ref{eq:keff} defines an experimentally measured Klyshko heralding efficiency for each of the three signal detectors. Each Klyshko heralding efficiency was measured for average powers between $P=10$ mW and $P=50$ mW at fiber lengths between $5$ m and $20$ m by using a single detector in the herald and signal channels. The data was fit by Eq.\ref{eq:nkpscale} with $\eta_{S}$, $\alpha$, and $\beta$ as free parameters. The Klyshko heralding efficiencies were similar for all three signal detectors. The data and fit are shown in Fig. \ref{fig:exkeff}.

	By fitting the measured Klyshko heralding efficiencies by Eq. \ref{eq:nkpscale}, we generate the dashed curve in Fig. \ref{fig:exkeff}. All four data sets are well fit by the same set of parameters: $\eta_{System} = 41.95\%$, $\beta = 5.324\times10^{-6}$ mW, and $\alpha=344$ mW-meters. The contribution due to Raman scattering is found to be negligible compared to the leaked pump noise for all fiber lengths measured. Therefore, additional measures such as cooling the fiber would not decrease the presence of noise in this experiment, although additional spectral and polarization filtering would decrease the pump photon contribution.

\begin{figure}[h!]
	\centering
	\includegraphics[width=1\textwidth]{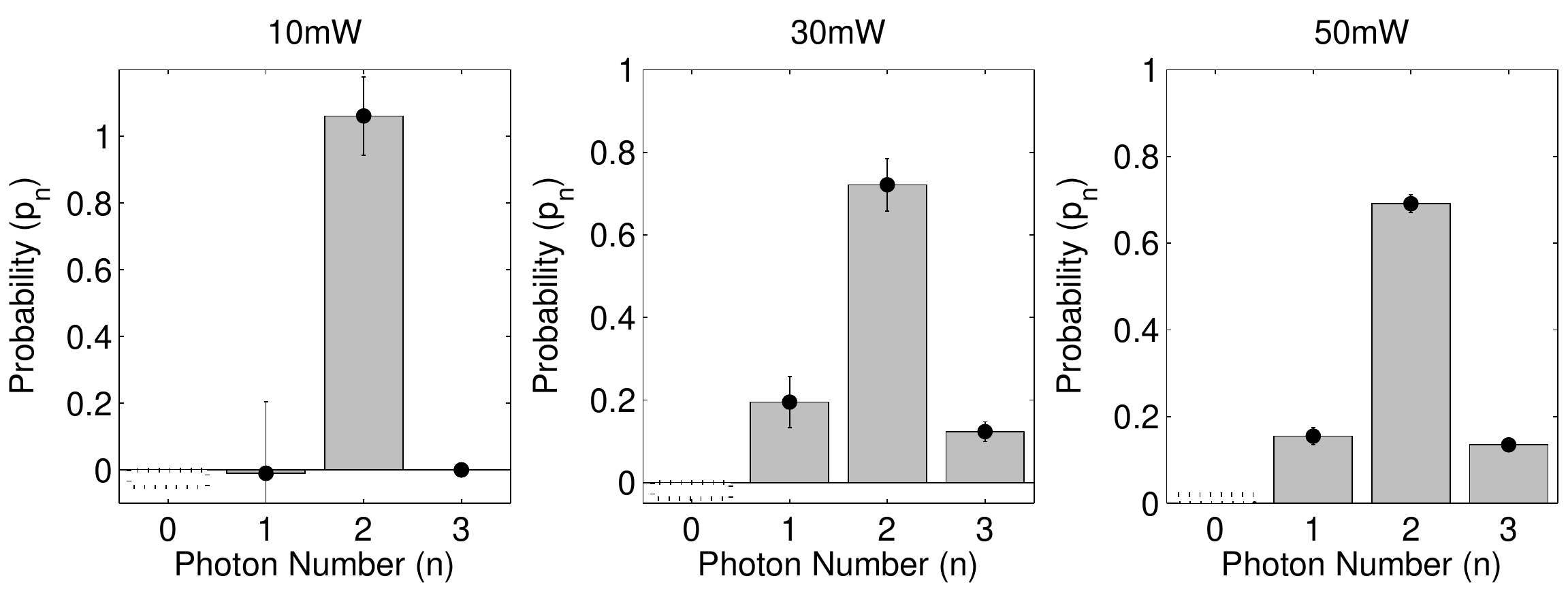}
	\caption{Inferred noise-free photon-number distributions with double-heralded conditional preparation using the Klyshko heralding efficiencies in the inversion algorithm. The measured values show the photon-number probabilities of the two-photon state that would be observed in the absence of false heralding from independent noise processes.}
	\label{fig:pnkeff}
\end{figure}

A major useful consequence of our noise analysis and measurement results is that from them we can calculate the effective photon-number distribution of the signal as would have been prepared in the absence of noise in the heralding channel. This is accomplished by modifying the inversion algorithm by substituting the experimentally measured, power-dependent Klyshko heralding efficiencies (see Fig. \ref{fig:exkeff}) in place of the tracer-beam-estimated signal arm efficiencies in  Eqs. \ref{eq:PAclick}-\ref{eq:PABCclick}. This procedure is valid if the extraneous photon scattering noise processes are independent of SFWM following the derivation of Eqs. \ref{eq:noiselesssig}-\ref{eq:pnkherald}. The resulting distributions, constructed from the same coincidence data as was used in Fig. \ref{fig:pn2hsyseff}, are presented in Fig. \ref{fig:pnkeff}. The surviving $p_1$ values are attributed to unwanted noise in the signal channel, which we make no attempt to remove here. The negative values of $p_0$ result because the linear inversion method does not constrain the values to be positive; they are within error bars (not shown) determined by those for the other $p_n$ values.

\section{Conclusion}
\label{sec:conclusion}
	We have studied double-heralded conditional preparation of states of light by spatially multiplexed detection in the presence of noise in the heralding channel. In contrast to experiments utilizing single-herald detection, we have shown the greater impact of noise photons in the herald channel on the photon-number statistics in double-heralded experiments. Furthermore, we have derived the equivalence between noise in the herald channel and loss in the signal channel. Through the temporal correlation of photon pairs, we were able to measure the efficiencies of detecting coincidences, the Klyshko heralding efficiencies, for each signal detector. We have demonstrated a method for inferring the signal channel photon-number distribution in the absence of heralding noise by using the Klyshko heralding efficiency in the conditional probability matrix. This method is useful as a diagnostic measurement to determine the relative contribution of noise in the herald channel from spontaneous Raman scattering and leaked pump photons for $\chi^{(3)}$ media. This diagnostic provides information regarding the necessity of additional precautions to mitigate noise sources in photon creation processes that are independent of the target process under measurement. 

\section{Acknowledgments}
	We are grateful for helpful discussions with Brian Smith, Ian Walmsley, and Eric Corwin. This work was supported by the NSF through the Physics QIS Program.  
\section*{Appendix: Conditional Probability Matrix Elements}
\label{sec:condprobapp}

	The elements of the conditional probability matrix in Eq. \ref{eq:pcmatrix} are:
\begin{align}
P(1|n)&=\mathscr{P}(A|n)+\mathscr{P}(B|n)+\mathscr{P}(C|n)-2\left(\mathscr{P}(AB|n)+\mathscr{P}(AC|n)+\mathscr{P}(BC|n)\right)\notag\\&+3\mathscr{P}(ABC|n)\label{eq:P1click}\\
P(2|n)&=\mathscr{P}(AB|n)+\mathscr{P}(AC|n)+\mathscr{P}(BC|n)-3\mathscr{P}(ABC|n)\label{eq:P2click}\\
P(3|n)&=\mathscr{P}(ABC|n)\label{eq:P3click}
\end{align}
	where $P(i|n)$ define the probability of $i$ and only $i$ detection events given $n$ incident photons and $\mathscr{P}(l|n)$ is the probability that all detectors in the set $l$ will click, given $n$ incident photons, where $l \in \{A,B,C,AB,...,ABC\}$. Eqs.\ref{eq:P1click}-\ref{eq:P3click} are calculated by invoking set theory to distinguish the one and only one, two and only two, or three detectors clicking. This method allows for the extraction of the number of detectors firing without recording which detectors clicked in each detection time bin. As a result, we are able to record data at high repetition rates and utilizing less memory.

	The probability of a set of detectors clicking takes into account the spatial multiplexing of the beam and the non-unity efficiency and the threshholding nature of the detectors. The probabilities are:
\begin{align}
\mathscr{P}(i|n)=&1-\left(1-R_i\eta_i\right)^n \label{eq:PAclick} \\
\mathscr{P}(ij|n)=&1-\left(1-R_i\eta_i\right)^n-\left(1-R_j\eta_j\right)^n+\left(1-R_i\eta_i-R_j\eta_j\right)^n\label{eq:PABclick} \\
\mathscr{P}(ijk|n)=&1-\left(1-R_i\eta_i\right)^n-\left(1-R_j\eta_j\right)^n-\left(1-R_k\eta_k\right)^n\notag\\
+&\left(1-R_i\eta_i-R_j\eta_j\right)^n+\left(1-R_i\eta_i-R_k\eta_k\right)^n+\left(1-R_j\eta_j-R_k\eta_k\right)^n\notag\\
-&\left(1-R_i\eta_i-R_j\eta_j-R_k\eta_k\right)^n\label{eq:PABCclick}
\end{align}
	where $i,j,k$ refer to detectors $\left\{A,B,C\right\}$, and $\eta_i$ is the total detection path efficiency for the $ith$ detector and $R_i$ is the fraction of light directed toward that detector. $\eta_i$ includes the transmission losses in the beam path and the quantum efficiency of the corresponding APD. For example, $(1-R_i\eta_i)^n$ is the probability that all $n$ photons are not detector at detector $i$. Thus $1-(1-R_i\eta_i)^n$ is the probability that at least one of the $n$ photons is detected at detector $i$, which is sufficient for threshold detection.
	
	It is clear from Eqs. \ref{eq:PAclick}-\ref{eq:PABCclick} that the conditional probability matrix is a function of the detection efficiency, including transmission losses in the optical path. The inversion reported in Section \ref{sec:results} was calculated using the measured optical transmission losses and estimated quantum efficiency of the APDs. The resulting inversion was the photon distribution in the signal channel with the addition of noise in the herald channel, producing false herald events. The inversion reported in Section \ref{sec:conclusion} was calculated with the Klyshko efficiencies in Eqs.  \ref{eq:PAclick}-\ref{eq:PABCclick}, providing the inferred, noise-free photon probability distribution, shown in Fig. \ref{fig:pnkeff}.

\end{document}